%% file: main.tex
\documentclass{article} %doublespace, 
\usepackage[utf8]{inputenc}
\usepackage{subfiles}

\usepackage{graphicx}
\usepackage{amsfonts}
\usepackage{hyperref}
\usepackage{amsmath}
\usepackage{multirow}
\usepackage{booktabs} 
\usepackage[skip=2pt plus1pt, indent=40pt]{parskip}
\usepackage{pdflscape}
\usepackage[doublespacing]{setspace}
\usepackage{geometry}
\title{Joint model with latent disease age: overcoming the need for reference time}
\author{Juliette ORTHOLAND$^1$, Nicolas GENSOLLEN$^1$, \\Stanley DURRLEMAN$^1$, Sophie TEZENAS DU MONTCEL$^1$\\[0.4em]  \small 1. Inria, centre of Paris, ARAMIS team, Sorbonne Universite,  Paris Brain \\ \small Institute AP-HP, INSERM, CNRS, University Hospital, Paris, France   \\ \small juliette.ortholand@icm-institute.org }

\date{}

\begin{document}

\maketitle

\noindent\textbf{Introduction} Heterogeneity of the progression of neurodegenerative diseases is one of the main challenges faced in developing effective therapies. Thanks to the increasing number of clinical databases, progression models have allowed a better understanding of this heterogeneity. Joint models have proven their effectiveness by combining longitudinal and survival data. Nevertheless, they require a reference time, which is ill-defined for neurodegenerative diseases, where biological underlying processes may start before the first symptoms.

\noindent\textbf{Objective} In this work, we propose a joint non-linear mixed effect model with a latent disease age, to overcome this need for a precise reference time.

\noindent\textbf{Method} To do so, we utilised an existing longitudinal model with a latent disease age as a longitudinal sub-model. We associated it with a survival sub-model that estimates a Weibull distribution from the latent disease age. We validated our model on simulated data and benchmarked it with a state-of-the-art joint model on real data from patients with Amyotrophic Lateral Sclerosis (ALS). Finally, we showed how the model could be used to describe ALS heterogeneity.

\noindent \textbf{Results} Our model got significantly better results than the state-of-the-art joint model for absolute bias (4.21(4.41) versus 4.24(4.14)(p-value=1.4e-17)), and mean cumulative AUC for right-censored events (0.67(0.07) versus 0.61(0.09)(p-value=1.7e-03)).

\noindent\textbf{Conclusion} We showed that our model is better suited than the state-of-the-art in the context of unreliable reference time, offering a new modelling approach. 

\paragraph{}

\newpage

\section{Introduction}
\par 
Neurodegenerative disorders are an important burden for the healthcare system (\cite{zahra_global_2020}). The heterogeneity of the progression of these diseases is a major challenge in the development of effective therapies, as in Alzheimer's Disease (\cite{duara_heterogeneity_2022}) or Amyotrophic Lateral Sclerosis (ALS) (\cite{beghi_heterogeneity_2007}). With the increasing number of large clinical databases, the development of disease progression models has helped in understanding this heterogeneity better. Two main data types can be subjected to modelling: longitudinal data such as repeated measures of clinical scores or biomarkers; or survival data, with the occurrences of events like death or surgical intervention. As for ALS , the U.S. Food and Drug Administration might demand assessing treatment efficacy using both types of outcomes (\cite{fdacderbastings_amyotrophic_2019}).

Data available for neurodegenerative disease is often sparse and covers only parts of the progression, which emphasises the need to realign different patients' ages depending on their disease stages to extract a full typical timeline of the disease (\cite{young_data-driven_2024}). The date of the first symptom is often used to do so, even if it does not capture the difference in terms of speed of progression. But in such disease the underlying biological processes start earlier than the first specific symptoms of the disease, making the first symptoms reported by the patient not representative of the disease onset and very subjective. For instance, in Alzheimer's Disease, the progression of amyloid and neurodegeneration biomarkers starts before the manifestation of the first clinical symptoms (\cite{jack_hypothetical_2010}). Similarly in ALS, changes in metabolism start before a significant weight loss and the first motor symptoms (\cite{peter_life_2017}), which again complicates the establishment of reliable reference time points for monitoring disease progression. Thus two challenges are faced to extract a full typical progression of these diseases: mapping the disease onset and the speed of progression of the patients.
\par 
Survival and longitudinal data are often associated with the same biological processes. In such a case, modelling both data together results in more precise estimates and improves inference (\cite{lu_two-level_2023}). Joint models enable us to do so. They are composed of three parts: a model for the longitudinal data, a model for the survival data and a linkage, often a shared latent variable, which captures the association between the two types of data.
\newline Classical survival models are hazard-based regression models (\cite{rubio_general_2019}). The most used is the Cox Proportional Hazard (Cox) model (\cite{cox_regression_1972}), with an effect on the hazard scale. Its main interests are that it does not require estimating the baseline hazard function when used alone and that it is easy to interpret due to the proportional impact of covariates on the baseline hazard. Nevertheless, the baseline hazard needs to be estimated for joint models (\cite{rizopoulos_joint_2012}) and the proportional assumption is often violated on long follow-up. Another family of models is the Accelerated Failure Time (AFT) model (\cite{kalbfleisch_statistical_2002}). In these models, covariates directly affect survival time, but modelling the hazard function is mandatory, making it less used for survival analysis alone. Besides, the results may be harder to interpret. %\textcolor{red}{Other approaches have enabled the description of the influence of unobserved covariates in a proportional hazards model using frailty models \cite{hougaard_frailty_1995}. Nevertheless, to do so, recurrent events or groups of patients should be available in the data. More recently, pseudo-observation techniques have enabled us to get rid of the censure problem in time-to-event data and enabled the treatment of these data as any repeated measure data \cite{andersen_pseudo-observations_2010}. However, this technique requires the restriction to the study of specific time points, which might not always be adapted. }
\newline
Classical longitudinal models are mixed-effect models that allow for modelling repeated and correlated observations (\cite{laird_random-effects_1982}). To capture the heterogeneity of the patient progression,  in addition to population parameters, named fixed effects, they have individual parameters, named random effects. The most common type of Non-Linear Mixed effects are generalised linear Mixed-effect Models (GLMMs), a direct extension of Linear mixed-effects models (LMM), which enables the estimation of non-linear progression (\cite{mcculloch_generalized_2008}). Such classical mixed-effect models rely on an empirical disease time axis which limits their temporal resolution to the resolution of the reference time used to index the disease time axis (\cite{young_data-driven_2024}). To overcome this issue, longitudinal models that capture the data-driven disease timeline behind the observable data, named data-driven disease progression models, have been developed( \cite{young_data-driven_2024}). Among others, disease Progression score models have been developed \cite{jedynak_computational_2012} and extended (\cite{bilgel_predicting_2019}) as well as models that allow direct integration of knowledge about different stages of diseases (\cite{raket_statistical_2020}). Finally, some models have explored the creation of a latent disease age (\cite{li_bayesian_2017, iddi_estimating_2018, schiratti_mixed-effects_2015, schiratti_methods_2017}). A non-linear-mixed effect model based on Riemannian geometry was proposed to capture a latent disease age and showed good performances in degenerative disease(\cite{schiratti_methods_2017,marinescu_tadpole_2019}). 
\newline
For the linkage structure between survival and longitudinal data, two types of modelling have been proposed: the latent class model (\cite{proust-lima_estimation_2017,proust-lima_describing_2023}) and the shared random effects model (\cite{rizopoulos_r_2016}). Such an approach aggregates similar patients which could help better understand the heterogeneity even though the meaning of the different classes remains unknown. It allows us to calculate the probability of an individual belonging to a particular class but may result in some patients being almost equitably distributed. Shared random effect models are often used to avoid these limits. Nevertheless, they have their own limitation: by including predictors of the longitudinal outcome in the survival model, they usually focus on how longitudinal outcomes impact survival. Whatever their type, all these joint models rely on GLMMs, necessitating the use of reliable reference times, which is not available in our context. 
\par
In this paper, we propose a latent age joint model suited for neurodegenerative diseases that overcomes the need for a reliable reference time of the state-of-the-art joint models. To do so, we used an existing longitudinal model with a latent disease age (\cite{schiratti_methods_2017}) as the longitudinal sub-model and used its defined latent disease age as the linkage structure. We associated a survival sub-model that estimates a Weibull distribution from the latent disease age. 
\newline
After introducing the proposed joint model, we validated it on simulated real-like clinical data. We then benchmark the proposed joint model against reference models on real ALS data and show that the proposed approach is better suited in the context of the absence of a reliable reference time point reference time than the state-of-the-art. Finally, we made a clinical application of the Joint Temporal model to present how parameters can be interpreted. 

\section{Model}\label{model_eq}
\subsection{Generic Framework}

\subsubsection{Data}

We consider $N$ patients, associated with longitudinal data: repeated measures of one given outcome $y$. Each patient $i$ is followed for $n_i$ visits. For each visit $j$, we denote $t_{i,j}$ the age at which the outcome is measured, and $y_{i,j}$ the value of the outcome for the patient at this visit.
\newline
We assume that we also observe an event $e$, and denote $t_{e_i}$ the age of the patient when the event is observed. Nevertheless, the event may occur after the follow-up period. In this case, the event is said to be censored, in opposition to observed. To distinguish censored and observed events, a boolean $B_{e_i}$ is associated with the time of the event $t_{e_i}$: $B_{e_i} = 0$ if the event is censored and $B_{e_i} = 1$ if the event is observed. If the event is censored, the time of the last visit is used as the time of the event (\cite{leung_censoring_1997}).

\subsubsection{Joint model structure}

The objective of joint models is to describe the combination of two types of clinical data: longitudinal data and survival data, with their relationship.  Here, the longitudinal process, $\gamma_i(t)$, is the progression of an outcome, measured by $y_{i,j}$ at each time $t_{i,j}$ for each visit $j$ of the patient $i$. The longitudinal process is estimated with a Gaussian noise $\epsilon \sim \mathcal{N}\left( 0, \sigma \right)$ compared to the measure, so that: $\nonumber y(t_{i,j}) = \gamma(t_{i,j})+\epsilon_{i,j}$. The survival process $S_i(t)$, the probability that a patient $i$ experiences the event after age $t$ ($S_i(t) = p(t_{e_i}>t)$). 
\newline
To describe the model, we will further use the formalism of mixed-effects models (\cite{laird_random-effects_1982}). Such models are composed of two types of parameters: parameters that differ from one patient to the other and enable to encapsulate the individual variability, named random effects, and parameters that capture the population specificity and are shared by all the patients, named fixed effects. 

%Note that $\gamma_0$ corresponds to the inverse of the link function of GLM used in Rizopoulos et al.\cite{rizopoulos_r_2016} and Proust-Lima et al. \cite{proust-lima_estimation_2017}. 

\subsection{The Proposed Joint model} \label{existing}

A non-linear mixed-effect model with a latent disease age was first introduced by \cite{schiratti_mixed-effects_2015}. Both the latent age and the modelling of the longitudinal process presented below are extracted from \cite{schiratti_mixed-effects_2015, schiratti_methods_2017}.

\subsubsection{The latent disease age: correction of individual variation}\label{time_rep}

The idea of the latent disease age, $\psi_i(t)$, is to map the chronological age of a patient into a latent disease age representative of the disease stage of the patient. Using the formalism described before, it can be written as :
\begin{eqnarray}
      \psi_i(t) =& e^{\xi_i}(t -\tau_i) + t_0
\end{eqnarray}
where $e^{\xi_i}$ is the progression rate of patient $i$, $\tau_i$ is its time-shift and $t_0$ is the population estimated reference time. 
\par
In the proposed joint model, the idea is to encapsulate all the individual variability of the patient $i$, in the latent disease age  $\psi_i$, with random effects. The latent disease age is then used as the link between the longitudinal and survival processes $(\gamma_i(t), S_i(t))$, which are estimated from the latent disease age with the composition of functions that describe only the population ($\gamma_0, S_0$) and are shared by all patients with fixed effects.  
\begin{eqnarray}
\begin{cases}
      \nonumber \gamma_i(t) = \gamma_0(\psi_i(t_{i,j}))\\
      S_i(t) = S_0(\psi_i(t))
    \end{cases}
\end{eqnarray}
For the survival modelling, instead of using time 0 as a start time, we use the reference time $t_0$ and impose $\forall t<t_0, S_0(t)=1$. Indeed, $t_0$ is automatically estimated thanks to the visit times and corresponds to a time of a given value of the score that most of the patients experimented with. Thus at that time, patients should not be dead. We have checked that even for the existing longitudinal model most patients were still alive at $t_0$. 
\newline
Compared to joint models with shared random effects models of (\cite{rizopoulos_r_2016}), where $S_0(\psi_i(t)) = f(\gamma_0(\psi_i(t))$ with $f$ a function of fixed parameters (further detailed in Table \ref{tab:ref_model}), the Joint Temporal model directly depends on the latent disease age.

\subsubsection{Modelling longitudinal process} \label{model_rm}
The modelling of the longitudinal process consists in computing the trajectory from the latent disease age defined in part \ref{time_rep}. We will study a clinical score with curvilinearity, and potential floor or ceiling effects (\cite{gordon_progression_2010}). Thus a logistic function will be used to model the outcome value from the latent disease age $\psi_i(t)$. It is parametrized as follows:
\begin{eqnarray}
      \gamma_0(\psi_i(t)) =& \left(1+g \times \exp(-{v_0}\frac{(g+1)^2}{g}(\psi_i(t)-t_0))\right)^{-1}
\end{eqnarray}
where $t_0$ is the population estimated reference time defined in \ref{time_rep}, $v_0$ is the speed of the logistic curve at $t_0$ and $\frac{1}{1+g}$ is the value of the logistic curve at $t_0$. To get the real value of the outcome $y_{i,j}$, the latent disease age $\psi_i(t)$ is first applied, then the longitudinal process from the latent disease age $\gamma_0 (t)$ and finally a Gaussian noise $\epsilon_{i,j}$ is added. We assume here that all the noises of visits are independent. The whole longitudinal process can thus be written as: 
\begin{eqnarray}
      y_{i,j} = \gamma_0(\psi_i(t_{i,j}))+\epsilon_{i,j} = \gamma_i(t_{i,j})+\epsilon_{i,j} \label{eq_long}
\end{eqnarray}
Note that using a Beta distribution for the noise instead of a Gaussian distribution would be more suited to a logistic trajectory. Nevertheless, in the context of small noise the Beta distribution could be approximated by a Gaussian distribution. %Such a hypothesis will be further tested in the simulations.  

\subsubsection{Modelling survival process}\label{model_tte}

A Weibull distribution is used to model the survival probability from the latent disease age $\psi_i(t)$:
\begin{eqnarray}
     \nonumber S_i(t) =&  S_0(\psi_i(t))= \mathbb{1}_{\psi_i(t) > t_0}\exp \left(- \left( \frac{\psi_i(t) - t_0}{\nu}\right)^{\rho}\right) + \mathbb{1}_{\psi_i(t) \leq t_0}
     %\nonumber =& \mathbb{1}_{t > \tau_i}\exp \left(- \left( \frac{e^{\xi_i}(t-\tau_i)}{\nu}\right)^{\rho}\right) + \mathbb{1}_{t\ \leq \tau_i}
\end{eqnarray}
where $\nu$ represents the variability of the distribution and $\rho$ the shape of the distribution. 
From there we also compute the individual hazard, which is, assuming that a patient has survived for a time t, the probability that he will not survive for an additional time $dt$:
\begin{eqnarray}
     \nonumber h_i(t) =& -\frac{S^{'}_i(t)}{S_i(t)} = \mathbb{1}_{\psi_i(t) > t_0} \frac{\rho e^{\xi_i}}{\nu}  \left(\frac{\psi_i(t) - t_0}{\nu}\right)^{\rho-1}
     %\nonumber =& \mathbb{1}_{t > \tau_i} \frac{\rho e^{\xi_i}}{\nu}  \left(\frac{e^{\xi_i} (t-\tau_i)}{\nu}\right)^{\rho-1}
\end{eqnarray}

\subsubsection{Joint Temporal model}\label{struct_model}
The proposed joint model referred to as the Joint Temporal model, is thus the combination of both a longitudinal sub-model $\gamma_i(t)$ and a survival sub-model $S_i(t)$ using the latent disease age $\psi_i(t)$ as a linkage structure, summarised in Table \ref{tab:ref_model}.
%
%\begin{eqnarray}
%\begin{cases}
%        \psi_i(t) = e^{\xi_i}(t -\tau_i) + t_0 \\
%       \gamma_i(t) = \gamma_0(\psi_i(t)) = \left(1+g \times \exp(-{v_0}\frac{(g+1)^2}{g}(\psi_i(t)-t_0))\right)^{-1}\\
%      S_i(t) = S_0(\psi_i(t))= \mathbb{1}_{\psi_i(t) > t_0}\exp \left(- \left( \frac{\psi_i(t) - t_0}{\nu}\right)^{\rho}\right) + \mathbb{1}_{\psi_i(t) \leq t_0}
%    \end{cases}
%\end{eqnarray}
%
\subsection{Estimation}

\subsubsection{Parameters}\label{param_structure}

For estimation purpose, latent parameters ($z$) are defined in addition to model parameters ($\theta$) and hyperparameters ($\Pi$). They can be summarised as follows for each patient $i$:

\begin{itemize}
\item Latent parameters ($z$):
    \begin{itemize}
        \item Latent fixed effects ($z_{fe}$): fixed effects sampled
        \begin{align*}
    \tilde{g} = \log(g) \sim \mathcal{N}\left( \overline{\tilde{g}}, \sigma^2_{\tilde{g}} \right) &&
    \tilde{v}_0 = \log(v_0) \sim \mathcal{N}\left( \overline{\tilde{v}_0}, \sigma^2_{\tilde{v}_0} \right)  &&\\
    \tilde{\nu} = -\log(\nu) \sim \mathcal{N}\left( \overline{\tilde{\nu}}, \sigma^2_{\tilde{\nu}} \right) &&
    \tilde{\rho} = \log(\rho) \sim \mathcal{N}\left( \overline{\tilde{\rho}}, \sigma^2_{\tilde{\rho}} \right)  &&
\end{align*}

        \item Latent random effects  ($z_{re}$): random effects sampled
        \begin{align*}
    \xi_i \sim \mathcal{N}\left( \overline{\xi}, \sigma^2_{\xi} \right) &&
    \tau_i \sim \mathcal{N}\left( \overline{\tau}, \sigma^2_{\tau} \right) &&
\end{align*}
    \end{itemize}
    \item Model parameters ($\theta$): fixed effects estimated from log-likelihood maximisation $\theta = \{ \sigma_{\xi}, \sigma_{\tau}, t_0, \overline{\tilde{g}}, \overline{\tilde{v}_0}, \overline{\tilde{\nu}}, \overline{\tilde{\rho}}, \sigma \}$
    \item Hyperparameters ($\Pi$): set by the user $\Pi = \{\sigma_{\tilde{g}}, \sigma_{\tilde{v_0}}, \sigma_{\tilde{\nu}}, \sigma_{\tilde{\rho}} \}$
\end{itemize}
To ensure identifiability, we set $\overline{\xi} = 0$ and $t_0 = \overline{\tau}$.

\subsubsection{Log-likelihood}

The likelihood estimated by the model is the following:
\begin{align*}
     \nonumber p(y,T_e, B_e \mid \theta, \Pi) =& \int_{z} p(y,T_e, B_e, z \mid  \theta, \Pi) dz
\end{align*}
$p(y,T_e, B_e, z \mid  \theta, \Pi)$ can be divided into two different terms: data attachment which represents how well the model describes the data $(y,t_e, B_e)$ and a prior attachment, which prevents over-fitting.
\begin{eqnarray}
     \log p((y,t_e, B_e), z,  \mid \theta, \Pi) =
    \nonumber & \log {p(y,t_e, B_e \mid z, \theta, \Pi)} + \log p(z\mid  \theta, \Pi )
\end{eqnarray}
The first term, data attachment, can be divided again into two terms considering that survival and longitudinal processes are independent regarding random effects. This is a quite common assumption in other papers (\cite{rizopoulos_joint_2012,proust-lima_estimation_2017}). We can also separate the prior attachment term: two terms for the prior attachment of latent parameters (fixed and random) and one term for the prior attachment of model parameters. We end up with the following expression : 
\begin{eqnarray}
     \nonumber \log p((y,t_e, B_e), z, \theta \mid \Pi) =& \log {p(y \mid z, \theta, \Pi)} + \log p(t_{e}, B_{e} \mid z, \theta, \Pi) \\
     \nonumber +& \log p(z_{re} \mid \theta, \Pi) + \log p(z_{fe} \mid \theta, \Pi) 
\end{eqnarray}
The different log-likelihood parts with their different assumptions and the total formula of the log-likelihood is available in appendix \ref{total_likelihood}.

\subsubsection{Algorithm}\label{cal_algo}

The first step is the estimation on the training dataset, it enables us to estimate fixed and associated random effects from a training data set. Directly maximising the log-likelihood has no analytical solution. Thus we use an Expectation-Maximization algorithm. Nevertheless, the computation of the expectation is also intractable due to the nonlinearity of the model. Thus, we use a Monte-Carlo Markov Chain Stochastic Approximation Expectation-Maximization (MCMC-SAEM) algorithm, as for the existing Longitudinal model. Its convergence has been proven by \cite{kuhn_coupling_2004} for models that lie in the curved exponential family. The Joint Temporal model falls into such a category and further details are given in appendix \ref{suf_stat}.  To get the mean of the distribution of the model, we apply a Robbins-Monro convergence algorithm to the last iterations  (\cite{robbins_stochastic_1951}). More details are given by \cite{koval_learning_2020} (p.41-43) and \cite{schiratti_methods_2017} (p.106). Latent parameters (defined in \ref{param_structure}), are estimated during the estimating phase of the EM algorithm and model parameters during the maximisation phase, using sufficient statistics. The total log-likelihood, the sufficient statistics and the maximisation update rules, necessary for the computation, are given in the appendices \ref{total_likelihood}, \ref{suf_stat} and \ref{max_rule}. 

The second step is the validation on a test set, to compute the random effects for new patients. During this step, the prediction of random effects for the patients is estimated using the standard approach by maximising the posterior distribution of the random effects given the visits and the censored event. The solver \textit{minimise} from the package Scipy (\cite{2020SciPy_NMeth}) was used to maximise the log-likelihood. Note that for predictions, the survival probability is then corrected using the survival probability at the last visit as in other packages (\cite{rizopoulos_r_2016}).
\newline
An implementation of the Joint Temporal model is available in the open-source library leaspy (v2):  \url{https://gitlab.com/icm-institute/aramislab/leaspy}.

\subsection{Reference models} 

\subsubsection{Reference models}\label{ref_model}

We chose to benchmark the Joint Temporal model against several reference models summurised with their equations in Table \ref{tab:ref_model}. First, we use one-process-only models. For the survival model, we use a Weibull AFT model to describe the survival process, using the Lifelines package (\cite{davidson-pilon_lifelines_2023}). This model will be referred to as the AFT model. For the longitudinal model, we use the existing Longitudinal model described in part \ref{existing} (Equation (\ref{eq_long})) using the open-source leaspy library  \url{https://gitlab.com/icm-institute/aramislab/leaspy}. This model will be referred to as the Longitudinal model. We expect the Joint Temporal model to be at least as good as these two models.
\newline
Second, we use a two-stage model: a survival model that uses the random effects of the Longitudinal model as covariates (\cite{murawska_two-stage_2012}). Even though this model is subject to an immortal bias, it enables us to compare our model to a better survival model than the AFT model. We use the Longitudinal model to extract random effects for each individual, and then use them as covariates in a Weibull AFT model, using the Lifelines package (\cite{davidson-pilon_lifelines_2023}). This model will be referred to as the Two-stage model, and the Joint Temporal model is expected to be at least as good as it. 
\newline
Third, we use a joint model with shared random effects, to evaluate if the newly proposed structure could improve estimation. To do so, we use a logistic longitudinal process, using the JMbayes2 package (\cite{rizopoulos_r_2016}). This model will be referred to as the JMbayes2 model.

\section{Simulation: model validation}

\subsection{Method}

For the simulation study we followed the ADEMP recommendation (\cite{morrisUsingSimulationStudies2019a}) and got inspiration from the work of (\cite{lavalley-morelleExtendingCodeOpensource2024}) also using MCMC-SAEM estimator.

\subsubsection{Aims} 

The simulation study aimed to validate the Joint Temporal model. For model parameters ($\theta$), we evaluate the estimation of the model parameters associated with their standard error. For random effects, we assessed the correlation between the estimated values and the simulated ones.

\subsubsection{Data-generatting mechanism}

Data were simulated under the Joint Temporal model structure with the following procedure:
\begin{enumerate}
    \item We simulated random effects using $\xi_i \sim \mathcal{N}\left( 0, \sigma^2_{\xi} \right)$ and $\tau_i \sim \mathcal{N}\left( t_0, \sigma^2_{\tau} \right)$.
\item We modelled the age at first visit $t_{i, 0}$ as $t_{i, 0} =  \tau_i + \delta_{f_i}$ with $\delta_{f_i} \sim \mathcal{N}\left( \overline{\delta_{f}}, \sigma^2_{\delta_{f}} \right)$.
\item We set a time of follow-up per patient $T_{f_i}$, with $T_{f_i} \sim \mathcal{N}\left( \overline{T_f}, \sigma^2_{T_f} \right)$ and a time between two visits $\delta_{v_i,j} \sim \mathcal{N}\left( \overline{\delta_{v}}, \sigma^2_{\delta_{v}} \right)$ to simulate $n_i$ visits until $t_{n_i} \leq t_{i, 0} + T_{f_i} < t_{n_{i+1}}$.
\item We set the value of the outcome at each visit using a beta distribution of concentration p and mode $\gamma_0(\psi_i(t_{i,j}))$ so that $y_{i,j} \sim \mathcal{B}\left( \gamma_0(\psi_i(t_{i,j})), p \right)$.
\item For each patient, we simulated the event $T_{e_i}$ through a Weibull distribution using $T_{e_i} \sim  e^{-\xi_i}\mathcal{W}\left( \nu , \rho \right) + \tau_i$.
\item  We considered that the event stopped the follow-up and that the follow-up censored the event. Thus all the visits after the event were censored: $t_{i,j}>T_{e_i}$ and events after the last visit were censored: $t_{i,max(j)}<T_{e_i}$.
\end{enumerate}

We simulated ALS real-like data using an ALS dataset, PRO-ACT, described in part \ref{real_data}, to get real-like values for parameters. Note that some parameters values were adjusted, such as the population estimated reference time, to make sure that no patient was left censored (Table \ref{table:simulation_param} in appendix). Parameters directly associated with the disease have been extracted from data analysis, using the Longitudinal and AFT models (Figures \ref{fig:ind_param}, \ref{fig:long_mod} in appendix). We simulated M=100 datasets with N=200 patients. The parameters used for the simulation study are summarised in Table \ref{table:simulation_param} in appendix.

\subsubsection{Estimands}

We initialised the Joint Temporal model with the Longitudinal model trained for 2,000 iterations and a survival Weibull model. Then, we ran the Joint Temporal model with 70,000 iterations (on average an hour) with the last 10,000 of the Robbins-Monro convergence phase (\cite{robbins_stochastic_1951}) to extract the mean of the posterior. 

We validated the estimation of the model parameters $\theta = \{ \sigma_{\xi}, \sigma_{\tau}, t_0, \overline{\tilde{g}}, \overline{\tilde{v}_0}, \overline{\tilde{\nu}}, \overline{\tilde{\rho}}, \sigma \}$ extracted by the Robbins-Monro convergence phase. As we use a Gaussian approximation for the noise, we estimated $\sigma$ using the noisy simulation and the expected perfect curve from the random effect used for simulation. For the random effects ($\tau_i, \xi_i$), to reduce the computation complexity, we extracted the mean of the last 10,000 iterations for each individual.

\subsubsection{Performance metrics}

To assess the estimation performances of the estimated model parameters ($\hat{\theta}$) over the M datasets simulated for the scenario, we reported:

\begin{itemize}
    \item the Relative Bias: $RB(\hat{\theta}) = \frac{1}{M} \sum_{m=1}^{M}\frac{\hat{\theta}^{(m)} - \theta}{\theta} \times 100$
    \item Relative Root Mean Square Errors: $RRMSE(\hat{\theta}) = \sqrt{\frac{1}{M}\sum_{m=1}^{M}\left(\frac{\hat{\theta}^{(m)} - \theta}{\theta} \times 100\right)^2}$
    \item Relative Estimation Errors: $REE^{(m)} = \frac{\hat{\theta}^{(m)} - \theta}{\theta} \times 100$
\end{itemize}
To assess the Standard Error of the estimated model parameters ($\hat{\theta}$), we reported:

\begin{itemize}
    \item the relative empirical Standard Error: $SE_{emp}(\hat{\theta}) = \sqrt{\frac{\sum_{m = 1}^M (\hat{\theta}^{(m)} - \Bar{\hat{\theta}})^2}{m-1}}$, $RSE_{emp}(\hat{\theta}) = \frac{SE_{emp}(\hat{\theta})}{\Bar{\hat{\theta}}}$
    \item the coverage rates (CR): defined as the proportion of datasets for which $\theta$ belonged to $[\hat{\theta}- 1.96 SE(\hat{\theta}),\hat{\theta} +1.96 SE(\hat{\theta})]$ with their 95\% confidence intervals (CI) computed using the exact Clopper Pearson method.
\end{itemize}
The estimation of the random effects ($\tau_i, \xi_i$) was assessed using the intraclass correlation between the mean of each individual and the true value that enabled the simulation.

\subsection{Results}

Over the 100 datasets, the censoring rate was around 81 (3)\% and the number of visits was in average of 2,104 (49) for the 200 patients. For a concentration parameter of 100 for the Beta distribution, the noise observed on the data was 0.04.

\begin{figure}[h]
\centering
    \includegraphics[width=0.5\linewidth]{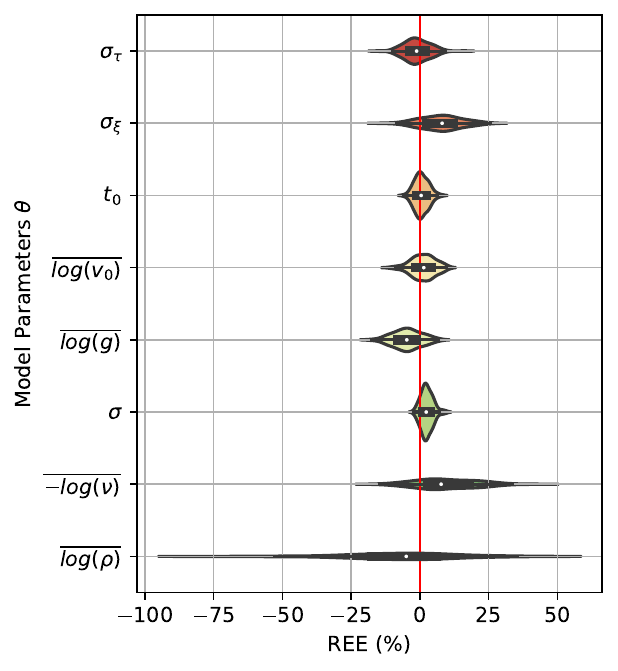}
    \caption{Relative Estimation Errors of estimated model parameters over 100 datasets } \label{fig:ree}
\end{figure}

The model got good estimation metrics (Table \ref{tab:metrics}) for the model parameters related to the random effects and the longitudinal fixed effects RRMSE below 11\%. For the survival fixed effects, the RRMSE was a bit higher (14.67 for the scale of the distribution ($\nu$) and 21.32 for the shape the distribution ($\rho$) which could also be explained by the difficulty of the set up small number of patients with a high censoring rate. The REE displayed in Figure \ref{fig:ree} confirms these results. The coverage rate CI did not include 95\% for half of the parameters but results were above 80\% on average.

Over the 100 datasets, the intraclass correlation for the estimated time reference ($\tau$) was of 0.96 (0.01) and of 0.92 (0.01) for the log speed factor ($\xi$).
\section{Application to ALS data: Variation in cohort description and prediction performance
benchmark}

\subsection{Method}

\subsubsection{ALS Data: PRO-ACT}\label{real_data}

We applied our method to data from an extraction of 2022 of the Pooled Resource Open-Access ALS Clinical Trials Consortium (PRO-ACT) database. It is composed of an aggregation of 23 phase II and III clinical trials and one observational study. It is a pseudonymised set of data with multiple inclusion and exclusion criteria for patients to enter the cohort (\cite{atassi_pro-act_2014}). For the study, we selected patients with at least three visits for the longitudinal outcomes described below (to ease prediction setup), age at first symptoms, symptom onset (spinal or bulbar) and sex.

As described by the FDA (\cite{fdacderbastings_amyotrophic_2019}), for ALS, clinical trials must demonstrate a treatment effect on function in daily activities and death. Following this guideline, we considered the most widely used functional rating system in patients with ALS, namely the revised version of the ALS functional rating scale revised (ALSFRSr) as longitudinal outcome (\cite{rooney_what_2017}). The scale starts at a maximum theoretical value of 48 and decreases with the severity of the disease till zero. For computation reasons, it was normalized using its scale so that the 0 value was the healthiest score and +1 the maximum disease score change. Tracheotomy was associated with death as a survival outcome, as in many ALS studies and challenges (\cite{kueffner_stratification_2019}).
\subsubsection{Prediction benchmark}

We compared the prediction of the reference models on real data, using a 10-fold cross-validation (90\% - 10\%). The Joint Temporal model was initialised with parameters of the Longitudinal model ran for 2,000 iterations and a Weibull model. It was further run with 70,000 iterations for each model (on average an hour and a half).  Personalisation was made using the first two visits of new patients and predictions were estimated on the remaining.

We wanted to evaluate both the goodness of survival and longitudinal predictions of the Joint Temporal model against reference models.
The goodness of longitudinal predictions was assessed using mean absolute error (MAE) and mean squared error (MSE), the latter being more sensible to outliers. The goodness of survival predictions was assessed using the C-index, for event order at 1 and 1.5 years, as in the ALS challenge (\cite{kueffner_stratification_2019}). Nevertheless, this method is not proper for evaluation (\cite{blanche_c-index_2019}), we thus added the mean cumulative dynamic AUC at 1 and 1.5 years for a correct measure but kept the C - index for comparison with existing results. We complement the metrics with an Integrated Brier Score (IBS), for the absolute distance between real and estimated events. The predictions were compared using a Wilcoxon signed-rank test with a Bonferroni adjustment for multiple pairwise comparisons. All the survival metrics were computed using the package sksurv (\cite{sksurv}).

\subsubsection{Cohort description}

The objective of this section was to demonstrate how the Joint Temporal model can be used to analyse death. After initialisation, we ran the Joint Temporal model on the whole PRO-ACT dataset presented in part \ref{real_data}, for 70,000 iterations (with a Robbins-Monroe convergence phase on the last 10,000 last iterations (\cite{robbins_stochastic_1951})). We extracted from the individual posteriors the mean of the random effects from 10,000 iterations. We used random effects to better characterize the heterogeneity associated with sex (man/woman) and onset site (spinal/bulbar). Differences between the four groups were assessed with ANOVA with Bonferroni correction. 

\subsection{Results}

\subsubsection{Data}
PRO-ACT datasets had similar or easier characteristics for estimation, compared to our real-like simulated dataset, in terms of the number of patients (2,528 compared to 200), visits (23,143 compared to 2,104 (49)) and censure rate (76.74 \% compared to 81 (3)\%) (Table \ref{tab:stat_proact} in appendix). 

\subsubsection{Prediction benchmark}

On real data, 18,077 longitudinal predictions were made on the remaining visits at an average time of 0.63 (0.55) years from the last visit.

For the longitudinal process, the Joint Temporal model had a significantly smaller MAE compared to JMbayes2 respectively 4.21 (4.41) and 4.24 (4.14) points of ALSFRSr (p-value = 1.4e-17), but a larger MAE compared to the Longitudinal model (4.18 (4.38) (p-value = 2.2e-73)) (Table \ref{tab:pred_real}).  The Joint Temporal model had a significantly higher MSE (37.11 (91.07)) compared to the JMbayes2 model (35.14 (71.33) (p-value = 1.1e-13)) and the Longitudinal model (36.67 (90.11) (p-value = 1.4e-63)). 
\newline
For the survival process, the Joint Temporal model was significantly better than all the other models for ordering events, with a mean AUC of 0.67 (0.07) (Table \ref{tab:pred_real}). The distance to the observed failure time was not significantly different from the one of the JMbayes2 model with an IBS of 0.1 (0.01), but it was significantly smaller than the one of the Two-stage model (0.11 (0.01) (p-value = 2.5e-04)) and the one of the AFT model (0.12 (0.01) (p-value = 1.1e-04)) (Table \ref{tab:pred_real}).

\subsubsection{Cohort description}

\begin{figure}[h]
\begin{center}
        \includegraphics[width=\linewidth]{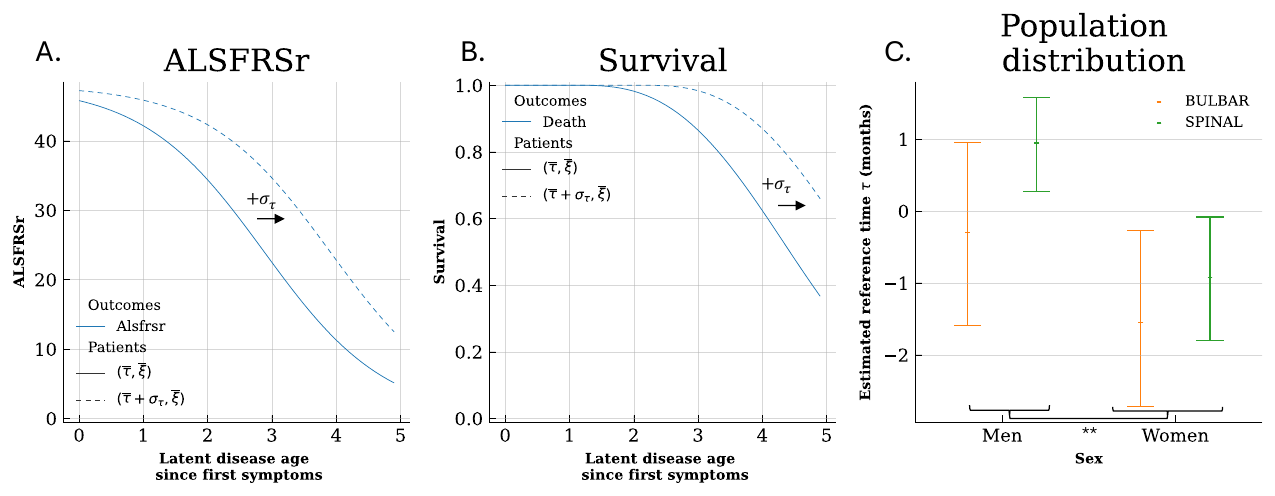}
           \caption{Individual estimated reference time and speed depending on sex and symptom onset } \label{fig:tau}
\end{center}
    \textit{\underline{Legend:} Panel A \& B: Impact of a positive variation of the estimated reference time ($\overline{\tau}+\sigma_{\tau}$) on the ALSFRSr progression (Panel A) and survival (Panel B) for the patient with average random effects ($\overline{\tau}$, $\overline{\xi}$),\\ 
    Panel C: Graph present the mean of random effects distribution for the four subgroups defined by sex (in abscissa men, women) and symptom onset (orange: Bulbar, green: Spinal) with its confidence interval 95\%. The vertical axis presents the estimated reference time in months compared to the mean estimated reference time of the whole population. ANOVA interaction p-value with Bonferroni correction: (A) 1. estimated reference time}
\end{figure}

\paragraph*{Estimated age at disease onset}
The estimated age at disease onset ($\tau$) impacts the start of the disease and shifts the curves of both ALSFRSr (Figure \ref{fig:tau} A) and survival (Figure \ref{fig:tau} B) without changing their shape. We did not find any significant interaction between the onset site and sex for the estimated age at disease onset (p-value = 1.) (Figure \ref{fig:tau} C). However, men were found to have a later estimated age at disease onset progress of 1.72 months (95\% CI = [0.78, 2.65]) than women independently of onset site.

\begin{figure}[h]
\begin{center}
    \includegraphics[width=\linewidth]{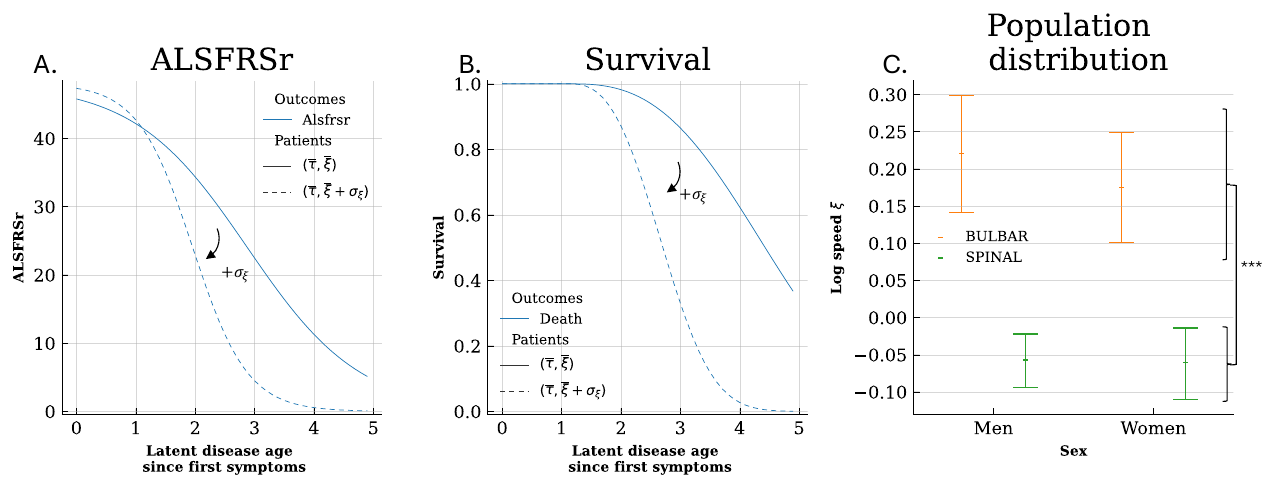}
   \caption{Individual estimated reference time and speed depending on sex and symptom onset } \label{fig:xi}
\end{center}
   \textit{\underline{Legend:} Panel A \& B: Impact of a positive variation of the estimated reference time  ($\overline{\xi}+\sigma_{\xi}$) on the ALSFRSr progression (Panel A) and survival (Panel B) for the patient with average random effects ($\overline{\tau}$, $\overline{\xi}$),\\ 
    Panel C: Graph present the mean of random effects distribution for the four subgroups defined by sex (in abscissa men, women) and symptom onset (orange: Bulbar, green: Spinal) with its confidence interval 95\%. The vertical axis presents the log speed compared to the mean log speed of the whole population. ANOVA interaction p-value with Bonferroni correction: 1. individual log-speed.}
\end{figure}

\paragraph*{Speed of progression}
The log speed ($\xi$) impacts the speed of the disease impacting the slope of the curves for both ALSFRSr (Figure \ref{fig:tau} A) and survival (Figure \ref{fig:tau} B). We did not find any significant interaction between the onset site and sex for the speed factor of progression (p-value = 1.) (Figure \ref{fig:xi} C). However, patients with bulbar onset were found to progress 1.29 times faster (95\% CI = [1.22, 1.38]) than patients with spinal onset independently of sex.

\section{Discussion}

Our work showed the potential of a latent disease age joint model to overcome the need for a precise reference time in neurodegenerative disease. 

Compared to the state-of-the-art it also alleviates the proportional hazard hypothesis while keeping an interpretable structure with the latent disease age. An underlying shared process is captured by the random effects and not a longitudinal-specific trend which offers a modelling alternative as illustrated in the application tasks. It has enabled us to realign the reference times, especially in terms of sex as the time references were found significantly different by f 1.72 months (95\% CI = [0.78, 2.65]), revealing that women might be more advanced at the time they report the first symptoms. The model also enabled to capture the individual speed of the disease extracted from both the survival and ALSFRSr which has confirmed that the disease progresses faster for bulbar onset compared to spinal onset (\cite{ortholandInteractionSexOnset2023, talbott_epidemiology_2016}). 

We showed how the Joint Temporal model could improve the prediction of ALS survival. Compared to the AFT model and the Two-stage model, the Joint Temporal model outperformed significantly all the metrics, but the Longitudinal model performed slightly better than the Joint Temporal model. Event censoring may not bring much to the Longitudinal model in this case, however, joint modelling corrected shortcomings of the two-stage model, such as immortal bias. Compared to the JMbayes2 model, the Joint Temporal model outperformed the longitudinal and event ordering metrics (C-index and AUC). This shows that our latent disease age might be better suited to capture the heterogeneity of the progression of degenerative diseases. Note that for the event distance metric (IBS) the JMbayes2 model and the Joint Temporal model did not perform significantly differently. This might be because the survival function of the JMbayes2 model, was more flexible, using splines instead of a Weibull function. Finally, we compared the C-index performances of the Joint Temporal model, with those of the ALS challenges that also used PRO-ACT data (\cite{kueffner_stratification_2019}). With a C-Index around 0.7 (0.5), we ranked seventh, even though we used only ALSFRSr progression to help the prediction, the other models being deep-learning models with multiple covariates. 
\newline
Longitudinal prediction got an MAE of 4.18 (4.38) points of ALSFRSr which is correct for a scale design on 48 points but that could still be improved compared to the Minimum Detectable Change (MDC) of 1.59 points of ALSFRSr(\cite{fournier_clinically_2023}).

Futur work on the longitudinal noise could be conducted including changing the distribution for a Beta distribution or alleviating the assumptions about the independence of longitudinal noises at each visit (\cite{proust-lima_estimation_2017}) could be done. More flexibility in the survival function could also be added to improve prediction performance. Integrating covariates would also facilitate the interpretation of the random effects distribution. Finally, modelling several longitudinal outcomes or events would enable covering a larger range of studies. 

In conclusion, the proposed joint model with latent disease age enabled us to improve the performance of most prediction metrics compared to existing joint models and alleviate the need for a precise reference time, offering a new modelling framework. This model opens up the perspective to design predictive and personalized therapeutic strategies.

\section{Acknowledgements}

Data used in the preparation of this article were obtained from the Pooled Resource Open-Access ALS Clinical Trials (PRO-ACT) Database. As such, the following organizations and individuals within the PRO-ACT Consortium contributed to the design and implementation of the PRO-ACT Database and/or provided data, but did not participate in the analysis of the data or the writing of this report: ALS Therapy Alliance, Cytokinetics, Inc., Amylyx Pharmaceuticals, Inc., Knopp Biosciences,  Neuraltus Pharmaceuticals, Inc., Neurological Clinical Research Institute, MGH, Northeast ALS Consortium, Novartis, Prize4Life Israel, Regeneron Pharmaceuticals, Inc., Sanofi, Teva Pharmaceutical Industries, Ltd., The ALS Association.

\section{Funding information}

The research leading to these results has received funding from the program "Investissements d’avenir" by the French government under management of Agence Nationale de la Recherche, reference ANR-10-IAIHU-06, ANR-19-P3IA-0001 (PRAIRIE 3IA Institute), ANR-19-JPW2-000 (E-DADS) and by H2020 programme reference 826421 (TVB-Cloud).

\section{Data availability statement}

The Joint Temporal model implementation is available on GitLab on the open-source package leaspy (v2): \url{https://gitlab.com/icm-institute/aramislab/leaspy}. The code for data simulation and experiments is available on a public gitlab repository: \url{https://gitlab.com/JulietteOrtholand/jm_article}. Note that to run all the experiments, access to a private library leaspype (on demand) and to the PRO-ACT dataset will be required.
\bibliographystyle{apalike}
\bibliography{WileyNJD-AMA}

\newpage

\begin{landscape}
\begin{table}[!ht]
\centering
\resizebox{1.2\textwidth}{!}{%
\begin{tabular}{ |c|c|c|c|c|c|c|c| } 
\hline
Model & Inputs &\multicolumn{2}{|c|}{Effects} & Random effects structure & \multicolumn{2}{|c|}{Link functions}  \\
&  &  Fix & Random & $\psi_i(t) $ & Longitudinal $\gamma(t)$ & Survival S(t)    \\
\hline
Longitudinal  & t & $g, v_0, t_0$ & $\xi, \tau $ & $e^{\xi}(t-\tau)+t_0$& $\left(1+g \times \exp(-{v_0}\frac{(g+1)^2}{g}(\psi_i(t)-t_0))\right)^{-1}$  & - \\ 

AFT & t & $\nu_0, \rho_0 $& - & - & - & $exp\left(-\left(\frac{t}{\exp(\nu_0)}\right)^{\exp(\rho_0)}\right)$ \\ 

Two-stages &$t, x = (\xi, \tau)$ & $\nu_0, \nu_1, \rho_0$ & - & - & - & $exp\left(-\left(\frac{t}{\exp(\nu_0 + \nu_1x)}\right)^{\exp(\rho_0)}\right)$ \\ 

Joint Temporal & t& $g, v_0, t_0, \rho, \nu$  & $\xi, \tau $ & $e^{\xi}(t-\tau)+t_0$& $\left(1+g \times \exp(-{v_0}\frac{(g+1)^2}{g}(\psi_i(t)-t_0))\right)^{-1}$  &  $exp \left(- \left( \frac{\psi_i(t)}{\nu}\right)^{\rho}\right)$ \\ 

JMBayes2 & t &$\beta_0, \beta_1, \alpha, spl(t)$ & $b_{i,0}, b_{i,1}$& $(\beta_0+b_{i,0}) + (\beta_1+b_{i,1})t$ & $\left(1+g \times \exp(\psi_i(t)\right))^{-1}$ &  $\exp \left( -\int_0^t \exp(spl(u)+ \alpha \gamma(u))du\right)$\\ 
\hline
\end{tabular}
}
\caption{Specification of reference models \\
\textit{\underline{Legend:} Longitudinal: existing longitudinal model with latent disease age, Two-stage model: AFT survival model that used as covariate random effects of the Longitudinal model, AFT: Accelerated Failure Time model, Joint Temporal: the proposed joint temporal model with latent disease age, JMbayes2:  joint model with shared random effects. $spl(t)$: spline function, $x = (\xi, \tau)$ are extracted from the Longitudinal model and used as covariates, }}
\label{tab:ref_model}
\end{table}

\end{landscape}

\begin{table}[]
    \centering
\caption{Performance metrics of the estimation of the model parameters of the Joint Temporal model\\ 
\textit{\underline{Legend:} RB: Relative Bias, RRMSE: Relative Root Mean Square Errors, CR: Coverage Rates, the metrics were computed over the 100 datasets simulated with 200 patients }} 
\begin{tabular}{ |c|lc|c|c|c| } 
\hline
 Type & Parameter name &  &        RB (\%) &      RRMSE (\%) & CR (\%) \\
 \hline
\multirow{2}{*}{Random Effect} & Time shift  (std)&$\sigma_{\tau}$ & -0.96 &   4.86 &  96.0 [90.1, 98.9]  \\ 
 & Individual log-rate factor  (std) &$\sigma_{\xi}$  &  7.65 &  10.69 &  83.0 [74.2, 89.8] \\ 
\hline
& {Population estimated reference time} &$t_0$  &  0.63 &   2.61 &  95.0 [88.7, 98.4] \\ 
Longitudinal & {Speed of the logistic curve} & $v_0$  &  1.25 &   4.42 &  93.0 [86.1, 97.1]  \\ 
Fixed Effects & {Curve value at $t_0$: $\frac{1}{1+g}$}& g  & -4.55 &   6.86 &  84.0 [75.3, 90.6]  \\ 
& Estimated noise &$\sigma$ &  2.60 &   3.41 &  81.0 [71.9, 88.2] \\
\hline
Survival & {Scale of the Weibull distribution} & $\nu$ &  9.86 &  14.67 &  84.0 [75.3, 90.6] \\ 
Fixed Effects & {Shape of the Weibull distribution} & $\rho$ & -6.55 &  21.32 &  94.0 [87.4, 97.8]  \\
\hline
\end{tabular}
\label{tab:metrics}
\end{table}

\begin{table}[!ht]
\centering
\caption{Prediction metrics of the Joint Temporal model and reference models on PRO-ACT data \\ 
\textit{\underline{Legend:} Joint Temporal: the proposed joint temporal model with latent disease age,  Two-stage model: AFT survival model that uses as covariate random effects of the Longitudinal model, Longitudinal: existing longitudinal model with latent disease age, AFT: Accelerated Failure Time model, JMbayes2:  joint model with shared random effects.\\
 Results are presented with the mean (SD) over the 10-fold cross-validation. P-values are computed using a Wilcoxon signed-rank test or pairwise t-test with Bonferroni between the Joint Temporal model and each of the reference models. IBS stand for Integrated Brier Score. \textdownarrow means that the metric should be minimised and \textuparrow maximised. Results in bold are the best for each metric. 18,077 longitudinal predictions were made at 0.63 (0.55) years from the last visit.}} 
\resizebox{\textwidth}{!}{%
\begin{tabular}{|l|l|ll|ll|ll|ll|}
\toprule
{} &         Joint Temporal&    Two stages &  p-value &  Longitudinal &  p-value &           AFT &  p-value &      JMbayes2 &   p-value \\
\hline
MAE \textdownarrow &   4.21 (4.41) &             - &        - &   \textbf{4.18 (4.38)} &  2.2e-73 &             - &        - &   4.24 (4.14) &  1.4e-17 \\
MSE    \textdownarrow       &  37.11 (91.07) &             - &        - &  36.67 (90.11) &  1.4e-63 &             - &        - &  \textbf{35.14 (71.33)} &  1.1e-13 \\
\hline
IBS \textdownarrow &   \textbf{0.10  (0.01)} &  0.11  (0.01) &  2.5e-04 &             - &        - &  0.12  (0.01) &  1.1e-04 &   0.10  (0.01) &  1.0e+00 \\
Mean AUC (1y, 1.5y)&  \textbf{0.67  (0.07)} &  0.62  (0.08) &  2.8e-03 &             - &        - &  0.42  (0.07) &  4.3e-05 &  0.61  (0.09) &  1.7e-03 \\
C-index 1.0y \textuparrow &  \textbf{0.69  (0.05)} &  0.63  (0.06) &  4.0e-04 &             - &        - &  0.41  (0.05) &  1.1e-06 &  0.63  (0.06) &  3.1e-04 \\
C-index 1.5y \textuparrow   &   \textbf{0.70  (0.05)} &  0.65  (0.05) &  9.7e-04 &             - &        - &  0.41  (0.05) &  1.6e-06 &  0.66  (0.05) &  1.8e-03 \\
\bottomrule
\end{tabular}}
\label{tab:pred_real}
\end{table}

\newpage
\vspace*{0.5cm}
\appendix
\subfile{9A_method}

\subfile{10A_simulation}

\end{document}

%% file: 9A_method.tex
\section{Likelihood total}\label{total_likelihood}
The likelihood estimated by the model is the following:
\begin{align*}
     \nonumber p(y,T_e, B_e \mid \theta, \Pi) =& \int_{z} p(y,T_e, B_e, z \mid  \theta, \Pi) dz
\end{align*}

and can be separated as follow:
\begin{eqnarray}
     \nonumber \log p((y,t_e, B_e), z, \theta \mid \Pi) =& \log {p(y \mid z, \theta, \Pi)} \\
     \nonumber +& \log p(t_{e}, B_{e} \mid z, \theta, \Pi) \\
     \nonumber +& \log p(z_{re} \mid \theta, \Pi) \\
     \nonumber +& \log p(z_{fe} \mid \theta, \Pi) 
\end{eqnarray}

The different log-likelihood parts are described below associated with their different assumptions. 

\paragraph{Longitudinal data attachment}\label{cal_data} For longitudinal process modelling, we assume that the longitudinal data are independent between patients and between visits given the random effects and that the noise of the process follows a Gaussian distribution. We thus get:
\begin{eqnarray}
       \nonumber \log {p(y \mid z, \theta, \Pi)} = \sum_{i,j} \log p(y_{i,j}\mid z,  \theta, \Pi) = \sum_{i,j} - \log\left(\sigma \sqrt{2\pi}\right) - \frac{1}{2\sigma^2} \left( y_{i,j} -  \gamma_0\left( \psi_i\left(t_{i,j}\right)\right)\right)^2
       \label{cal_rm_eq}
\end{eqnarray}
\paragraph{Survival data attachment} For survival process modelling, we once again assume that all patients are independent and that the modelling of the survival process depends on whether the event is observed or not:
\begin{eqnarray}
     \nonumber \log p(t_{e}, B_{e} \mid z, \theta, \Pi) =& \sum_{i}\log p(t_{e_i}, B_{e_i} \mid z, \theta, \Pi) \\
     \nonumber =& \sum_{i} \mathbb{1}_{B_{e_i}} \times \log \left(h_i(t_{e_i})\right) + \sum_{i} \log \left(S_i(t_{e_i})\right)
\end{eqnarray}
Note that if $\psi_i(t) < t_0$ $\log \left(h_i(t_{e_i})\right)=-\infty$. To prevent estimation issues, we initialise the algorithm at a possible point getting inspiration from barrier methods \cite{nesterov_lectures_2018}.

\paragraph{Latent random effects priors attachment } As patients are supposed to be independent of each other, we suppose that random effects are independent conditionally to $\theta$ and $\Pi$. The regularization term associated, with $\overline{\xi} = 0$, is then:
\begin{eqnarray}
    \nonumber \log p(z_{re} \mid \theta, \Pi)
    =& {\log p(\tau_i \mid \theta, \Pi)}
         + {\log p( \xi_i \mid \theta, \Pi)}\\
    \nonumber =& -{N \log\left( \sigma_{\tau} \sqrt{2\pi} \right) - \frac{1}{2\sigma^2_{\tau}}\sum_i (\tau_i - t_0)^2} \\
    \nonumber  &-{N \log\left( \sigma_{\xi} \sqrt{2\pi} \right) - \frac{1}{2\sigma^2_{\xi}}\sum_i (\xi_i - \overline{\xi})^2}
\end{eqnarray}

\paragraph{Latent fixed effects prior attachment } Each latent fixed effect is independently sampled from a posterior distribution. The regularization term associated is then:
\begin{eqnarray}
    \log p(z_{fe} \mid \theta, \Pi)
    \nonumber =& {\log p(\tilde{g} \mid \theta, \Pi)} + {\log p(\tilde{v_0} \mid \theta, \Pi)} \\
     \nonumber + &{\log p(\tilde{\nu} \mid \theta, \Pi)} + {\log p(\tilde{\rho} \mid \theta, \Pi)}\\
    \nonumber= &-  {\log\left( \sigma_{\tilde{g}} \sqrt{2\pi} \right) - \frac{1}{2\sigma^2_{\tilde{g}}} \left( \tilde{g} - \overline{\tilde{g}} \right)^2}\\
   \nonumber  - &{\log\left( \sigma_{\tilde{v}_0} \sqrt{2\pi} \right) - \frac{1}{2\sigma^2_{\tilde{v}_0}} \left( \tilde{v}_0 - \overline{\tilde{v}_0} \right)^2}\\
    \nonumber - &{\log\left( \sigma_{\tilde{\nu}} \sqrt{2\pi} \right) - \frac{1}{2\sigma^2_{\tilde{\nu}}} \left( \tilde{\nu} - \overline{\tilde{\nu}} \right)^2}\\
    \nonumber -&{\log\left( \sigma_{\tilde{\rho}} \sqrt{2\pi} \right) - \frac{1}{2\sigma^2_{\tilde{\rho}}} \left( \tilde{\rho} - \overline{\tilde{\rho}} \right)^2}
\end{eqnarray}

\begin{align*}
     \nonumber & &&\log p((y,T_e, B_e), z \mid \theta, \Pi) \\
    \nonumber &=&& \sum_{i,j} - \log\left(\sigma \sqrt{2\pi}\right) - \frac{1}{2\sigma^2} \left( y_{i,j} -  \gamma_0\left( \psi_i\left(t_{i,j}\right)\right)\right)^2 \\
    \nonumber    &+&& \sum_{i} \mathbb{1}_{B_{e_i}} \times \log \left(h_i(t_{e_i})\right) + \sum_{i} \log \left(S_i(t_{e_i})\right)\\
    \nonumber & - && {\log\left( \sigma_{\tilde{g}} \sqrt{2\pi} \right) - \frac{1}{2\sigma^2_{\tilde{g}}} \left( \tilde{g} - \overline{\tilde{g}} \right)^2}\\
   \nonumber & - &&{\log\left( \sigma_{\tilde{v}_0} \sqrt{2\pi} \right) - \frac{1}{2\sigma^2_{\tilde{v}_0}} \left( \tilde{v}_0 - \overline{\tilde{v}_0} \right)^2}\\
    \nonumber& -&&{\log\left( \sigma_{\tilde{\nu}} \sqrt{2\pi} \right) - \frac{1}{2\sigma^2_{\tilde{\nu}}} \left( \tilde{\nu} - \overline{\tilde{\nu}} \right)^2}\\
    & -&&{\log\left( \sigma_{\tilde{\rho}} \sqrt{2\pi} \right) - \frac{1}{2\sigma^2_{\tilde{\rho}}} \left( \tilde{\rho} - \overline{\tilde{\rho}} \right)^2}\\
    \nonumber &-&&{N \log\left( \sigma_{\tau} \sqrt{2\pi} \right) - \frac{1}{2\sigma^2_{\tau}}\sum_i (\tau_i - t_0)^2} \\
    \nonumber &-&&{N \log\left( \sigma_{\xi} \sqrt{2\pi} \right) - \frac{1}{2\sigma^2_{\xi}}\sum_i (\xi_i - \overline{\xi})^2} 
       \label{cal_tot_likelihood}
\end{align*}

\section{Sufficient statistics}\label{suf_stat}

The convergence of the Monte-Carlo Markov Chain Stochastic Approximation Expectation-Maximization (MCMC-SAEM) algorithm has been proven in \cite{kuhn_coupling_2004} for models which lie into the curved exponential family. For such a family of distributions, the log-likelihood can be written as:
\begin{equation}\label{exp_family}
    \nonumber \log \ p(\textbf{Y, z}, \theta, \Pi) = - \Phi(\theta, \Pi) + \langle S(\textbf{Y, z}), f(\theta, \Pi)\rangle
\end{equation}
where $\Phi$ and $f$ are smooth functions, and $S$ are called the sufficient statistics. The sufficient statistics are to be understood as a summary of the required information from the latent variables $\textbf{z}$ and the observations $\textbf{Y}$ in our case $(y, t_e, B)$. Our model falls in such a category and sufficient statistics are described below. 

The idea is to rewrite likelihood in the above form to get sufficient statistics. As a reminder, note that there are $N$ patients indexed by $i$ and each has $n_i$ visits indexed by $j$.

\begin{equation*}
    \begin{split}
        \log q((y,T_e, B_e),z\mid \theta ,\Pi) = & \sum_{i,j} - \log\left(\sigma \sqrt{2\pi}\right) - \langle \underbrace{[\| y_{ij} \|^2]_{ij}}_{S_1(Y, z)} \underbrace{- 2 [y_{ij}^T \gamma_0\left( \psi_i\left(t_{i,j}\right)\right)]_{ij}}_{S_2(Y, z)} + \underbrace{[\| \gamma_0\left( \psi_i\left(t_{i,j}\right)\right) \|^2]_{ij}}_{S_3(Y, z)} , \frac{1}{2\sigma^2} \mathbf{1}_{\sum n_i} \rangle\\
        & + \langle \underbrace{[\mathbb{1}_{B_{e_i}} \times \log \left(h_i(t_{e_i})\right) + \sum_{i} \log \left(S_i(t_{e_i})\right)]_i}_{S_4(Y, z)}, \mathbf{1}_{N} \rangle \\
        & -  \ln(\sigma_{\tilde{g}} \sqrt{2 \pi}) + \langle \underbrace{[\tilde{g}^2]}_{S_5(Y, z)}, -\frac{1}{2\sigma^2_{\tilde{g}}} \mathbf{1}_{1} \rangle + \langle \underbrace{[\tilde{g}]}_{S_6(Y, z)}, \frac{1}{\sigma_{\tilde{g}}^2}  [\overline{\tilde{g}}] \rangle - \sum \limits_{k=1}^{1} \frac{1}{2 \sigma_{\tilde{g}}^2} \overline{\tilde{g}}^2 \\
        & -  \ln(\sigma_{\tilde{v}_0} \sqrt{2 \pi}) + \langle \underbrace{[\tilde{v}_0^2]}_{S_7(Y, z)}, -\frac{1}{2\sigma^2_{\tilde{v}_0}} \mathbf{1}_{1} \rangle + \langle \underbrace{[\tilde{v}_0]}_{S_8(Y, z)}, \frac{1}{\sigma_{\tilde{v}_0}^2}  [\overline{\tilde{v}_0}] \rangle - \sum \limits_{k=1}^{1} \frac{1}{2 \sigma_{\tilde{v}_0}^2} \overline{\tilde{v}_0}^2 \\
        & -  \ln(\sigma_{\tilde{\nu}} \sqrt{2 \pi}) + \langle \underbrace{[\tilde{\nu}^2]}_{S_9(Y, z)}, -\frac{1}{2\sigma^2_{\tilde{\nu}}} \mathbf{1}_{1} \rangle + \langle \underbrace{[\tilde{\nu}]}_{S_{10}(Y, z)}, \frac{1}{\sigma_{\tilde{\nu}}^2}  [\overline{\tilde{\nu}}] \rangle - \sum \limits_{k=1}^{1} \frac{1}{2 \sigma_{\tilde{\nu}}^2} \overline{\tilde{\nu}}^2 \\
        & -  \ln(\sigma_{\tilde{\rho}} \sqrt{2 \pi}) + \langle \underbrace{[\tilde{\rho}^2]}_{S_{11}(Y, z)}, -\frac{1}{2\sigma^2_{\tilde{\rho}}} \mathbf{1}_{1} \rangle + \langle \underbrace{[\tilde{\rho}]}_{S_{12}(Y, z)}, \frac{1}{\sigma_{\tilde{\rho}}^2}  [\overline{\tilde{\rho}}] \rangle - \sum \limits_{k=1}^{1} \frac{1}{2 \sigma_{\tilde{\rho}}^2} \overline{\tilde{\rho}}^2 \\
        & - {N} \log (\sigma_\tau \sqrt{2 \pi}) + \langle \underbrace{[\tau_i^2]_i}_{S_{13}(Y, z)} , -\frac{1}{2 \sigma_\tau^2} \mathbf{1}_{N} \rangle  + \langle \underbrace{[\tau_i]_i}_{S_{14}(Y, z)} , \frac{1}{\sigma_\tau^2} \overline{\tau} \mathbf{1}_{N} \rangle - \frac{1}{2\sigma_\tau^2} {N} \overline{\tau}^2 \\
        & - N \log (\sigma_\xi \sqrt{2 \pi}) + \langle \underbrace{[\xi_i^2]_i}_{S_{15}(Y, z)} , -\frac{1}{2 \sigma_\xi^2} \mathbf{1}_{N} \rangle  + \langle \underbrace{[\xi_i]_i}_{S_{16}(Y, z)} , \frac{1}{\sigma_\xi^2} \overline{\xi} \mathbf{1}_{N} \rangle - \frac{1}{2\sigma_\xi^2} N \overline{\xi}^2 \\
    \end{split}
\end{equation*}

\section{Maximization update rules}\label{max_rule}

To find the update rule of the different parameters, we need to find the new parameter $\theta$ that maximizes the log-likelihood. As expressions are convex in $\theta$ we can simply derive and look for a critical point. We derive the log-likelihood with respect to each maximised fixed effect. Note that only maximised fixed effects are updated by a maximization rule, other parameters are latent variables that are sampled. $\overline{\xi}$ is first maximised and then set to 0. As a reminder, note that there are $N$ patients indexed by $i$ and that each of them has $n_i$ visits indexed by $j$. At iteration $k$, we can use $\tilde{S}^{(k+1)}$ computed with the parameters at iteration k and the formula of $S(Y, z)$ to compute the parameters at iteration $(k + 1)$.

\begin{align*}
    (\sigma^2)^{(k+1)} & \leftarrow \frac{1}{N} [\tilde{S}^{(k)}_1 - 2 \tilde{S}^{(k)}_2 + \tilde{S}^{(k)}_3]^T \mathbf{1}_1\\
    (\overline{\tilde{g}}_j)^{(k+1)} & \leftarrow \tilde{S}^{(k)}_6\\
    (\overline{\tilde{v}_0}_j)^{(k+1)} & \leftarrow \tilde{S}^{(k)}_8\\
    (\overline{\tilde{\nu}}_j)^{(k+1)} & \leftarrow \tilde{S}^{(k)}_{10}\\
    (\overline{\tilde{\rho}}_j)^{(k+1)} & \leftarrow \tilde{S}^{(k)}_{12}\\
    (\overline{\tau})^{(k+1)} & \leftarrow
    \frac{1}{N}\tilde{S}^{(k)}_{14}\\
    (\sigma^2_{\tau})^{(k+1)} & \leftarrow \frac{1}{p} [\tilde{S}^{(k)}_{13} - 2 \overline{\tau} \tilde{S}^{(k)}_{14}]^T \mathbf{1}_{N} + \overline{\tau}^2\\
    (\overline{\xi})^{(k+1)} & \leftarrow
    \frac{1}{N}\tilde{S}^{(k)}_{16}\\
    (\sigma^2_{\xi})^{(k+1)} & \leftarrow \frac{1}{N} [\tilde{S}^{(k)}_{15} - 2 \overline{\xi} \tilde{S}^{(k)}_{16}]^T \mathbf{1}_{N} + \overline{\xi}^2
\end{align*}

%% file: 10A_simulation.tex
\section{Analysis of PRO-ACT for real-like simulation}

\begin{figure}[!ht]
    \centering
    \includegraphics[width=0.8\linewidth]{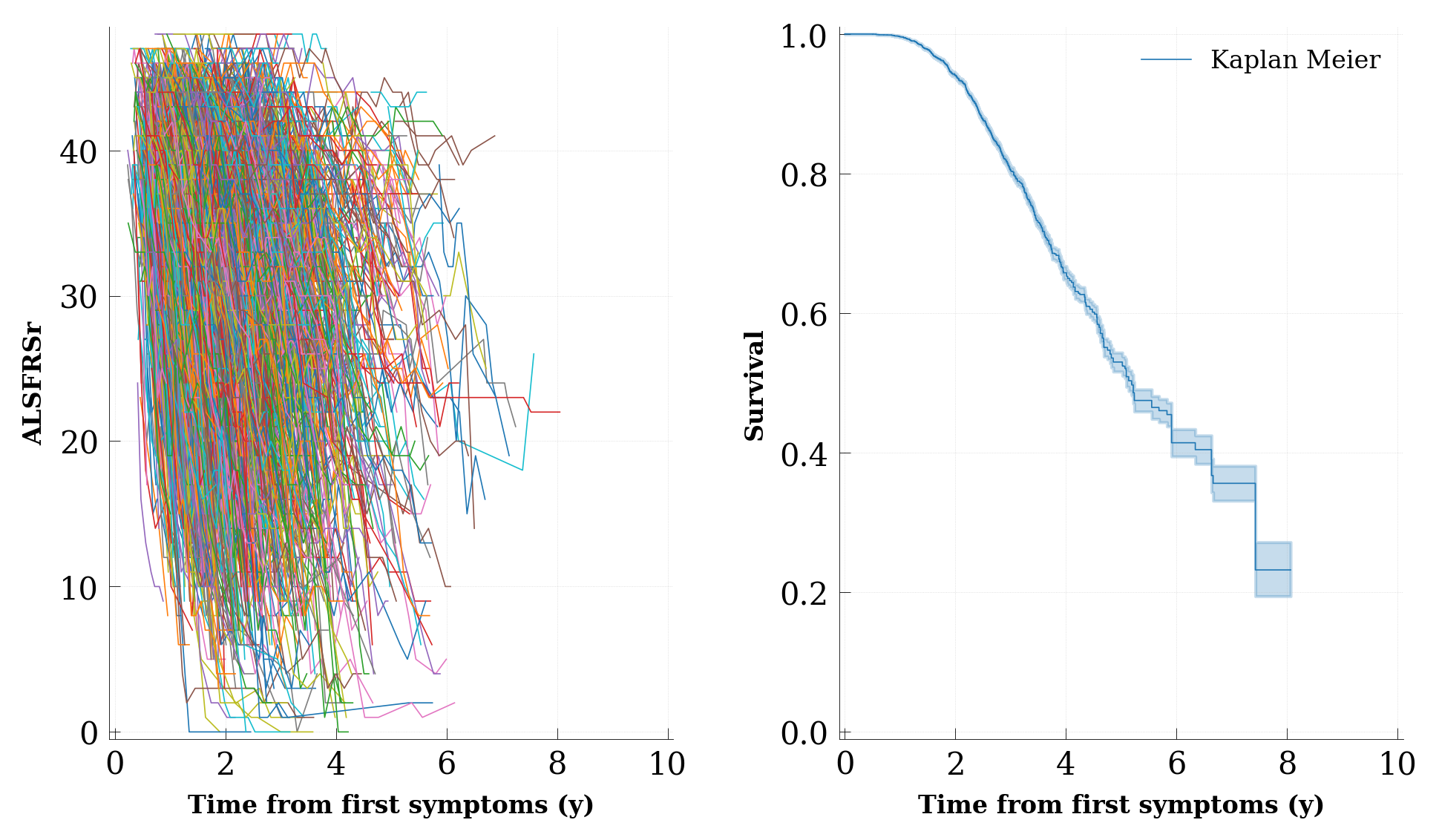}
\caption{PRO-ACT longitudinal and survival data \\ \textit{\underline{Legend:} The spaghetti plot represents each patient trajectory of ALSFRSr from first symptoms. Survival corresponds to time to death or tracheotomy.}} 
    \label{fig:ind_param}
\end{figure}

\begin{figure}[!ht]
    \centering
    \includegraphics[width=0.8\linewidth]{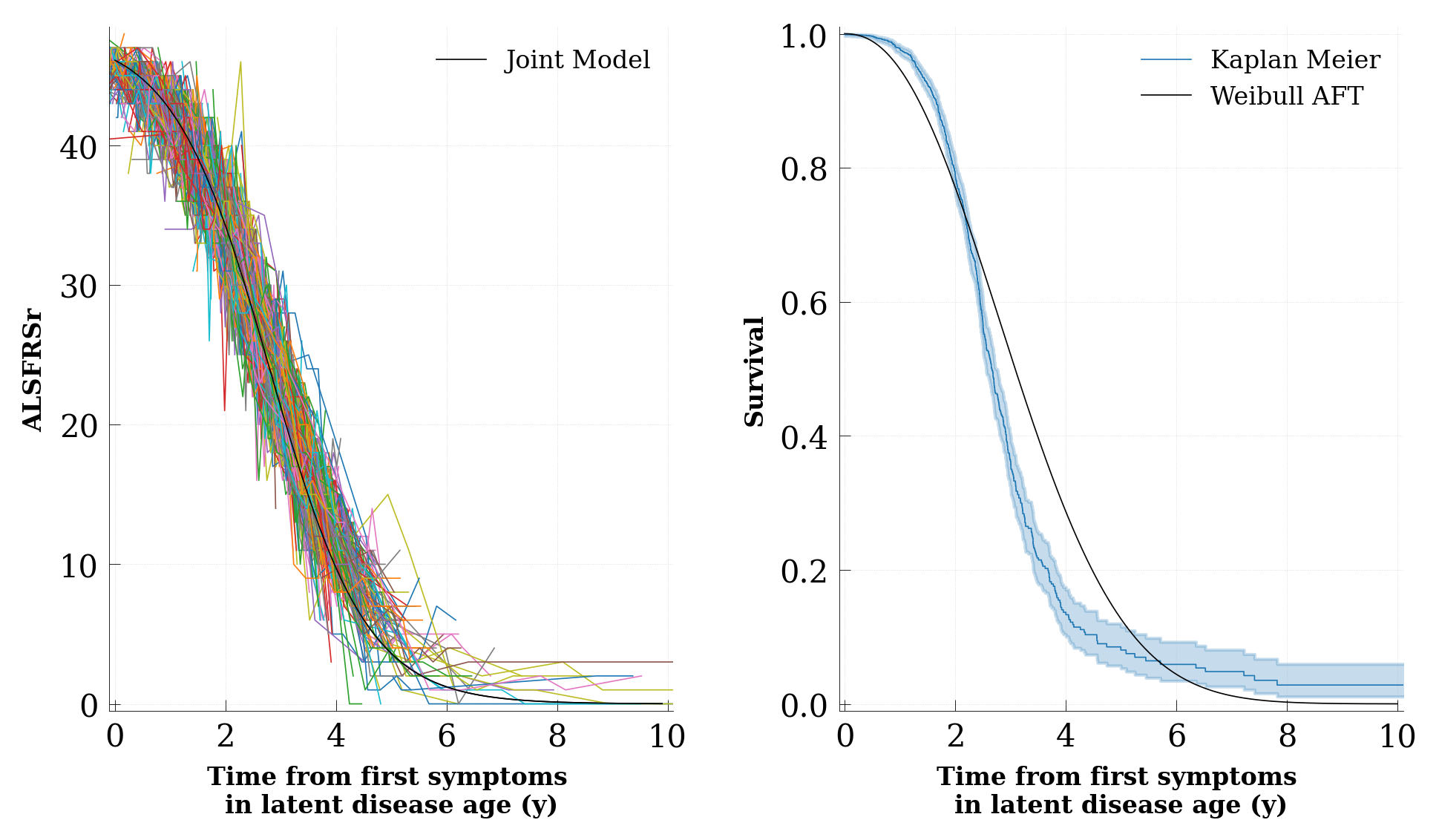}
\caption{PRO-ACT longitudinal and survival data reparametrized in latent disease age\\ \textit{\underline{Legend:} The spaghetti plot represents each patient trajectory of ALSFRSr from first symptoms in latent disease age. Survival corresponds to time to death or tracheotomy in latent disease age.}} 
    \label{fig:long_mod}
\end{figure}

\section{Summary of parameters for simulation}

\begin{table}[!ht]
\begin{center}
\centering
\caption{Parameters used for data simulation\\ \textit{\underline{Legend:} (r) indicate when ALS real-like parameters are used. Note that the population estimated reference time, $t_0$, corresponds also to the mean time shift.}}
\begin{tabular}{ |c|cc|c|c|c| } 
\hline
 Type & \multicolumn{3}{|c|}{Parameters}  & ALS & Used for \\
  & \multicolumn{2}{|c|}{Name} & Symbol & real-like (r) & simulation \\
 \hline
 Patients & {Patient number} &&$N$ & 2,528 & 200 \\ 
\hline
\multirow{3}{*}{Random Effect} & Time shift & (std)&$\sigma_{\tau}$ & 1.04 & r \\ 
 & \multirow{2}{*}{Individual log-rate factor} &(mean) &$\overline{\xi}$ & 0 & r \\ 
& & (std) &$\sigma_{\xi}$ & 0.73 & r \\ 
\hline
& {Population estimated reference time}& &$t_0$ & 1.17 & 5  \\ 
Longitudinal & {Speed of the logistic curve} && $v_0$ & 1.13 & r  \\ 
Fixed Effects & {Curve value at $t_0$: $\frac{1}{1+g}$}&& g & 6.40 & r  \\ 
& Estimated noise &&$\sigma$ & $\mathcal{N}(y, 0.04)$ & $\mathcal{B}(y, 100)$\\
\hline
Survival & {Scale of the Weibull distribution} && $\nu$ & 3.62 & r  \\ 
Fixed Effects & {Shape of the Weibull distribution} && $\rho$ & 2.25 & r  \\
\hline
\multirow{5}{*}{Visits}& \multirow{2}{*}{Time between $\tau$ and baseline} & (mean) & $\overline{\delta_{f}}$ & 0.4  & 0 \\ 
& & (std) &$\sigma_{\delta_{f}}$ & 0.84 & 0.4 \\ 
& \multirow{2}{*}{Time of follow up} &(mean)&$\overline{T_f}$ & 0.96& 1.2 \\ 
&&(std)&$\sigma_{T_f}$ & 0.87 & 0.3  \\ 
& \multirow{2}{*}{Time between visits} &(mean)&$\overline{\delta_{v}}$ & 1.47 & r \\ 
& &(std)&$\sigma_{\delta_{v}}$ & 0.5  & r \\ 
\hline
\end{tabular}
\label{table:simulation_param}
\end{center}
\end{table}

\section{Real data description}

\begin{table}[!ht]
\centering
\caption{Characteristics of the PRO-ACT data \\ \textit{\underline{Legend:} Results are presented with mean (SD) [class\%]. There were no missing values in the dataset due to patient selection.}} 
\begin{tabular}{|c|c|}
\toprule
Characteristics                    &         Values            \\
\hline
Number of patients                       &              2,528 \\
Number of visits                         &             23,143 \\
Number of patients years                 &          2,531 \\
Percentage of censored events (\%)        &         76.74 \\
Number of visits per patients                 &         9.2 (4.3) \\
Time of follow-up (years)                    &         1.0 (0.6) \\
Time between visits (months)                  &  1.5 (0.9) \\
\hline
Gender (Male)                            &     1,575 [62.3 \%] \\
Symptom onset (Spinal)                   &     1,952 [77.2 \%] \\
Age at first symptoms                    &       54.0 (11.3) \\
Time from first symptoms to baseline (years) &         1.6 (0.9) \\
ALSFRSr total (baseline)                       &        37.9 (5.4) \\
\bottomrule
\end{tabular}
\label{tab:stat_proact}
\end{table}

\section{Benchmark - real data} 

\begin{figure}[!ht]
    \centering
    \includegraphics[width=\linewidth]{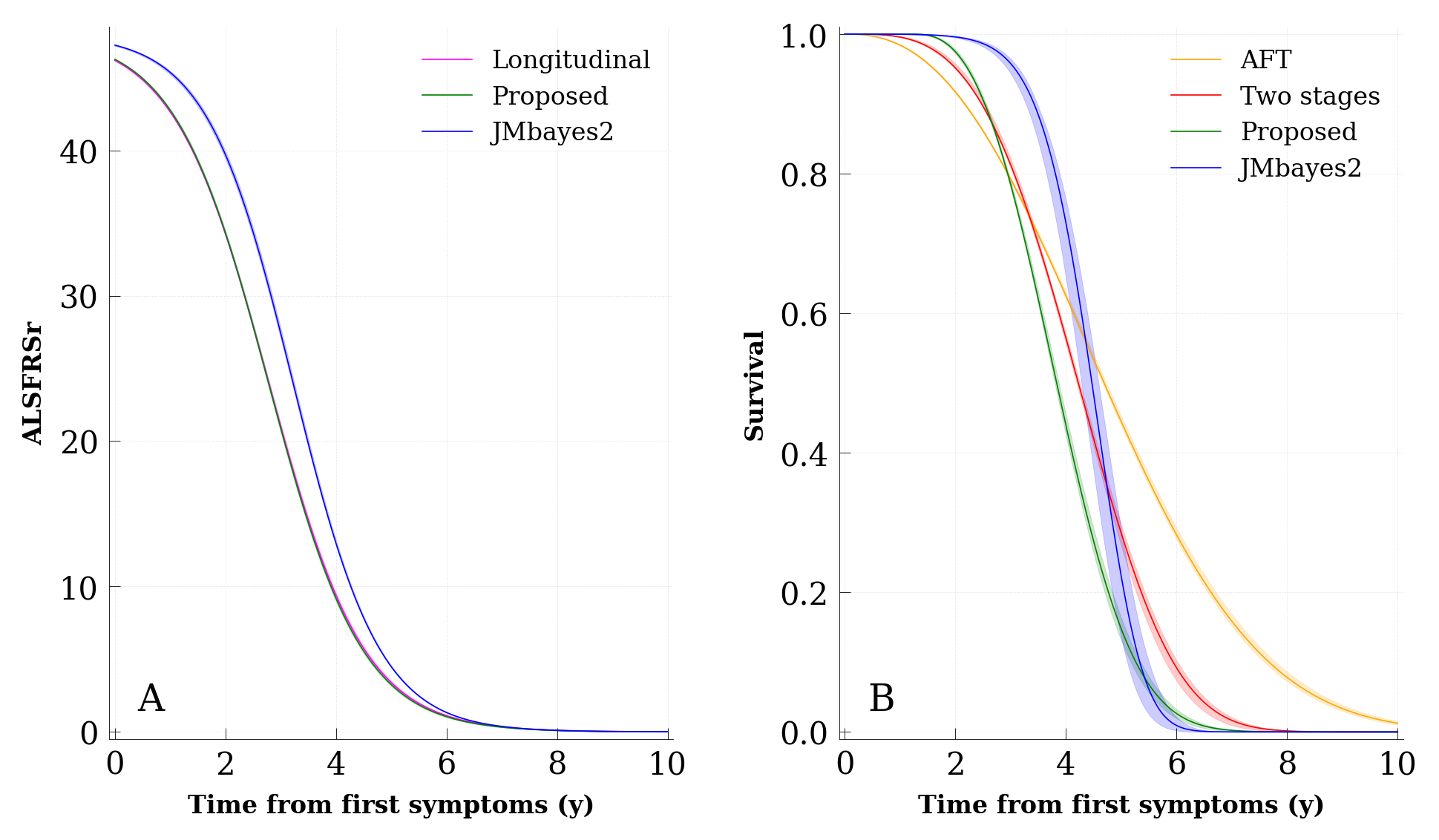}
    \caption{Average patient curve of the Proposed model and reference models on PRO-ACT dataset \\ \textit{\underline{Legend:}  Results are presented with mean over the 10-fold cross-validation and the maximum and minimum variation for each model.\\
    \textbf{Panel A:} Longitudinal: existing longitudinal model with latent disease age, Proposed: the proposed joint temporal model with latent disease age, JMbayes2:  joint model with shared random effects, \\
    \textbf{Panel B:}  AFT: Accelerated Failure Time model, Two-stage model: AFT survival model that uses as covariate random effects of the Longitudinal model, Proposed: the Proposed model, JMbayes2:  joint model with shared random effects.\\
    Note that for the longitudinal process, the curves of the Longitudinal and Proposed models are superimposed.}} 
    \label{fig:average_pat}
\end{figure}